# Palaeosymbiosis revealed by genomic fossils of Wolbachia in a strongyloidean nematode

Georgios Koutsovoulos[1*], Benjamin Makepeace[2], Vincent N. Tanya[3], Mark Blaxter[1]

1 Institute of Evolutionary Biology, The University of Edinburgh, Edinburgh EH9 3JT, UK

2 Institute of Infection and Global Health, The University of Liverpool, Liverpool L3 5RF, UK

3 Institut de Recherche Agricole pour le Développement, Regional Centre of Wakwa, Ngaoundéré, BP65 Adamawa Region, Cameroon

| | |
|---|---|
| Georgios Koutsovoulos | g.d.koutsovoulos@sms.ed.ac.uk |
| Benjamin Makepeace | blm1@liverpool.ac.uk |
| Vincent Tanya | vntanya@yahoo.com |
| Mark Blaxter | mark.blaxter@ed.ac.uk |

* author for correspondence:

Georgios Koutsovoulos, Institute of Evolutionary Biology, The Ashworth Laboratories, The King's Buildings, The University of Edinburgh, Edinburgh EH9 3JT, UK

g.d.koutsovoulos@sms.ed.ac.uk

Competing interests: The authors declare no competing interests.

# Abstract

*Abstract*

*Wolbachia* are common endosymbionts of terrestrial arthropods, and are also found in nematodes: the animal-parasitic filaria, and the plant-parasite *Radopholus similis*. Lateral transfer of *Wolbachia* DNA to the host genome is common. We generated a draft genome sequence for the strongyloidean nematode parasite *Dictyocaulus viviparus*, the cattle lungworm. In the assembly, we identified nearly 1 Mb of sequence with similarity to *Wolbachia*. The fragments were unlikely to derive from a live *Wolbachia* infection: most were short, and the genes were disabled through inactivating mutations. Many fragments were co-assembled with definitively nematode-derived sequence. We found limited evidence of expression of the *Wolbachia*-derived genes. The *D. viviparus Wolbachia* genes were most similar to filarial strains, and strains from the host-promiscuous clade F. We conclude that *D. viviparus* was infected by *Wolbachia* in the past. Genome sequence based surveys are a powerful tool for revealing the genome archaeology of infection and symbiosis.

*Introduction*

*Wolbachia* are alphaproteobacterial, intracellular symbionts of many non-vertebrate animal species, related to rickettsia-like intracellular pathogens such as *Anaplasma* and *Ehrlichia* (Werren, 1997). *Wolbachia* was first detected as a cytoplasmic genetic element causing mating type incompatibilities in *Culex pipiens* mosquitoes, and subsequently has been found to infect many insect species (Zug and Hammerstein, 2012). In insects, most *Wolbachia* can be classified as reproductive parasites, as they manipulate their hosts' reproduction to promote their own transmission (Werren et al., 2008). This is achieved by induction of mating type incompatibilities, induction of parthenogenesis in females of haplo-diploid species, and killing or feminisation of genetic males. In some insects, *Wolbachia* infections are apparently "asymptomatic", in that no reproductive bias has been detected. There is evidence that *Wolbachia* infection can be beneficial to hosts, particularly in protection from other infectious organisms (Walker et al., 2011). Importantly, in most insect systems tested the symbiosis is not essential to the hosts, which can be cured by antibiotic treatment.

*Wolbachia* strains have been classified into a number of groups using molecular phylogenetic analyses of a small number of marker loci (Lo et al., 2007, Comandatore et al., 2013). Insect *Wolbachia* largely derive from clade A and B. Outside Insecta, arthropod *Wolbachia* infections have been identified in terrestrial Collembola (Hexapoda), Isopoda (Crustacea), Chelicerata and Myriapoda, and also in marine Amphipoda and Cirripeda (Crustacea). Most non-insect arthropod infections also involve *Wolbachia* placed in clades A or B. A minority of arthropod infections involves *Wolbachia* placed in distinct lineages (clades E through N) (Lo et al., 2007, Augustinos et al., 2011). In clade A and B symbionts, transmission appears to be essentially vertical (mother to offspring) in ecological time, but phylogenetic analysis reveals that horizontal transfer between hosts has been common on longer timescales.

*Wolbachia* infections have also been identified in nematodes, notably in the animal parasites of the Onchocercidae. These filarial parasites utilise arthropod vectors (dipterans and chelicerates) in transitioning between their definitive vertebrate hosts, but the *Wolbachia* they carry are not closely related to those of the vector arthropods. The majority of *Wolbachia* from onchocercid nematodes are placed in two distinct but related clades, C and D (Bandi et al., 1998, Comandatore et al., 2013). The biology of the interaction between filarial nematodes and their C and D *Wolbachia* is strikingly different (Foster et al., 2005). There is no evidence of reproductive manipulation. Transmission is vertical, as in other *Wolbachia*, but, unlike the arthropod symbionts, in species with infections all members carry the symbionts, and the phylogeny of hosts and symbionts show remarkable congruence. Treatment with antibiotics both kills onchocercid nematode *Wolbachia*, and also affects the viability of the nematodes, suggesting a strongly mutualistic, possibly essential interaction (Genchi et al., 1998, Hoerauf et al., 1999). The interaction is not essential on a phylogenetic timescale, as nested within the *Wolbachia*-infected onchocercids are species that have lost their infections (McGarry et al., 2003). The biological bases for the mutualism is a topic of significant research interest, and may include manipulation of embryogenesis, metabolic provisioning and modulation of host immune responses (Taylor et al., 2000, Foster et al., 2005, Fenn and Blaxter, 2004, Fenn and Blaxter, 2006, Darby et al., 2012).

Not all nematode *Wolbachia* are placed in clades C and D (Casiraghi et al., 2005). Clade F *Wolbachia* have a distinct host profile compared to the other clades, as they have been found in both onchocercid nematodes (*Mansonella*, *Madathamugadia* and *Cercopithifilaria* species) (Lefoulon et al., 2012), and arthropods (hexapods and chelicerates). The *Wolbachia* symbiont from the nematode *Dipetalonema gracile* is the sole representative of clade J, but is closely related to clade C *Wolbachia* (Casiraghi et al., 2004). A *Wolbachia* infection has been described in *Radopholous similis*, a tylenchid plant parasitic nematode distantly related to the Onchocercidae (Haegeman et al., 2009). This symbiont has been placed in a new clade I. The biological role(s) of these nematode *Wolbachia* have yet to be defined. *Wolbachia* have been sought in other nematode species, both parasitic and free-living. These searches, carried out using *Wolbachia*-specific PCR amplification of marker genes, have generally proved negative in individuals sampled across the diversity of Nematoda other than Onchocercidae (Bordenstein et al., 2003). In the many ongoing nematode genome and transcriptome projects, *Wolbachia*-derived sequence has only been described from onchocercid nematodes and *R. similis*. However, there are two overlapping expressed sequence tags from *Ancylostoma caninum* (also a member of Strongyloidea) that have high similarity to *Wolbachia* genes (Elsworth et al., 2011), but these have not been verified as derived from a *Wolbachia* symbiont in this species. (The relationships of the nematode taxa discussed are illustrated in Figure 1 (Blaxter, 2011, Lefoulon et al., 2012, Ferri et al., 2011).)

Horizontal transfer of *Wolbachia* genome fragments into the host nuclear genome has been detected in arthropods and nematodes that carry live infections (Blaxter, 2007, Fenn et al., 2006). Inserted fragments range from what is likely the whole bacterial genome inserted into an azuki beetle chromosome, to short fragments at the limit of specific detection. These fragments have excited much debate, particularly concerning the Onchocercidae, where it has been hypothesised that they may represent functional gene transfers into the nematode genome and thus play significant roles in host biology (Dunning Hotopp et al., 2007, McNulty et al., 2012, McNulty et al., 2013, Ioannidis et al., 2013). However most *Wolbachia* insertions have accumulated many substitutions and insertion-deletion events compared to their functional homologues in extant bacterial genomes. In this they most resemble nuclear insertions of mitochondrial DNA, which are 'dead on arrival' and evolve neutrally in the host chromosome (Blaxter, 2007).

Interestingly, the onchocercid nematodes *Onchocerca flexuosa* (McNulty et al., 2010), *Acanthocheilonema viteae* (Casiraghi et al., 2001, Hoerauf et al., 1999) and *Loa loa* (McGarry et al., 2003) lack *Wolbachia* despite their placement within the group of *Wolbachia*-containing species. This suggests that they have lost their live *Wolbachia* infections. Fragments of *Wolbachia*-like sequence have been detected in the nuclear genome in these species (McNulty et al., 2010, Desjardins et al., 2013). *Wolbachia* nuclear transfers, or *nuwt*s (Blaxter, 2007), in nematodes that currently lack live *Wolbachia* infection can be thought of molecular fossils of the previous symbiosis history of the host. Just as fossil skeletal remains can reveal the past distribution of larger biota, *nuwt*s can reveal past symbioses, and their divergence from current *Wolbachia* genomes can be used to estimate the date of the symbiosis.

We are engaged in a phylum-wide survey of genomes within the Nematoda (Kumar et al., 2012). As part of our analytic procedures we routinely screen raw genomic DNA data for contamination with environmental, commensal and host DNAs with a pipeline that uses read coverage, contig GC% and sequence identity to known protein sequences (Kumar et al., 2013). This serves to identify, and ease removal of, contaminating genomes, which in turn improves target genome assembly and aids independent assembly of symbiont genomes where present. Here we present an analysis of genome sequence data from the strongyloidean nematode *Dictyocaulus viviparus*, the bovine lungworm, which reveals molecular fossils of an ancient *Wolbachia* symbiosis in this economically important species, which is only distantly related to the previously known nematode hosts (Figure 1).

*Results*

**The draft genome sequence of *Dictyocaulus viviparus***

We generated a draft genome for the strongyloidean nematode *D. vivparus* based on a single adult male specimen provided from a cow slaughtered at an abattoir in Ngaoundéré, Cameroon. The *D. viviparus* genome was assembled using Velvet from 16 gigabases of cleaned data from 165 million, 100-base, paired-end reads from a 500 base pair (bp) insert library sequenced on an Illumina HiSeq2500 instrument. The draft assembly spanned 169.4 megabases (Mb) (Table 1). In terms of contiguity, the draft was of moderate quality with an N50 (length of contig at which 50% of the genome is in contigs of this size or larger) of 22 kilobases (kb), and N90 of 5 kb. There were 17,715 contigs above 500 bp. The assembly had a GC content of 34.5% and estimated read coverage of ~80 fold (Figure 2A). The mitochondrial contigs from the assembly had >99.5% identity to the published mitochondrial genome of *D. viviparus*. The size of this draft assembly is within the range of published genome sizes from species of the same suborder (Rhabtitina), which range from 80 Mb (*Heterorhabditis bacteriophora* (Bai et al., 2013)) to 320 Mb (*Haemochus contortus* (Bai et al., 2013)) (Table 2). Given that we used a single library, and had no long-range mapping data, it is likely that this genome size estimate is lower than the true genome as near-identical repeats will have been collapsed or left unassembled. We assessed the completeness of the draft assembly using the Core Eukaryotic Genes Mapping Approach (CEGMA (Parra et al., 2007)), and identified 90% complete and 93% partial genes. A previous Roche 454 transcriptome assembly for *D. viviparus* (Cantacessi et al., 2011) was used to assess the assembly's completeness in terms of representation of known *D. viviparus* transcripts. Retaining matches where over 70% of the transcript were mapped to the same genome contig, 87% of transcripts were present in the assembly. Many additional transcripts were split between contigs.

Using a MAKER2-Augustus pipeline (Cantarel et al., 2008, Stanke et al., 2006), we predicted 14,306 protein-coding genes, with a median length of 834 bp, median exon length of 168 bp, and a median of 7 exons per gene. We compared this predicted gene set for *D. viviparus* to those of *Caenorhabditis elegans* (The *C. elegans* Genome Sequencing Consortium, 1998), *H. bacteriophora* (Bai et al., 2013) and *H. contortus* (Schwarz et al., 2013, Laing et al., 2013) using orthoMCL (Li et al., 2003). A majority (75%) of the predicted *D. viviparus* proteins clustered with proteins from these rhabditine nematodes (Figure 2). The only species which had a low proportion of proteins clustered was *H. bacteriophora* (~40%), an observation that has been noted previously (Bai et al., 2013).

Thus, while the goal of our study was not to produce a high-quality reference genome for *D. viviparus*, the draft assembly and annotation produced are still of reasonable quality (Table 2). A majority of known *D. viviparus* genes are present, similarity to related nematode species is high, and most of the genes appear to be present and in full length. The genome assembly and a dedicated BADGER genome exploration environment (Elsworth et al., 2013) are available from http://dictyocaulus.nematod.es.

**Identification of *Wolbachia*-like sequences in the nuclear assembly**

As part of our standard quality control processes, we generated a taxon-annotated GC-coverage plot (TAGC plot) (Kumar et al., 2013), with the goal of identifying any non-nematode (either bovine host or environmental bacterial) contamination (Figure 3 A). This process allows identification of contaminants by their presence as contigs with differing GC content or estimated read coverage compared to that of assured target genome contigs (Kumar and Blaxter, 2011). The taxonomic annotation, using the NCBI BLAST+ suite, serves to assign contaminant contigs to their possible species of origin. This process identified a total of 193 contigs, spanning 1 Mb, that had best matches to *Wolbachia* (Figure 3 B). The *Wolbachia*-like contigs had a GC content very close to the mode for the nematode genome, but they had a wide range of estimated coverages, from approximately equal to the majority of nematode-derived contigs to 3-4 fold higher Figure 3 C). Unusually, the *Wolbachia*-like contigs were not better assembled than the nuclear genome. The lower complexity of the alphaproteobacterial genome usually results in more contiguous assembly, even at low coverage.

The putative *Wolbachia* from *D. viviparus* (*w*Dv) contigs were compared to the complete genomes of *Wolbachia* from *Brugia malayi* (*w*Bm) (Foster et al., 2005) and *O. ochengi* (*w*Oo) (Darby et al., 2012). The average identity of the BLAST hits was 84.5% ±3.2% to both of the other *Wolbachia* genomes, indicating similar evolutionary distance from these two taxa (Figure 3 D). The matches were distributed across the genomes of other *Wolbachia* (Figure 3 B). The *Wolbachia*-like fragments were uploaded to the RAST server (Aziz et al., 2008) for direct annotation, and 1580 coding sequences were predicted, almost double than found in previous nematode *Wolbachia* genomes (http://rast.nmpdr.org/?page=JobDetails&job=112231; Table 3). This elevated number largely resulted from frameshifts and stop codons in the middle of genes, which fragmented the open reading frames, and overall only 567 different *Wolbachia* genes (of a usual 800 to 1500) were identified. We also screened the contigs that had *Wolbachia* matches for other informative similarities, and identified 29 that contained both nematode and *Wolbachia* matches (examples are illustrated in Figure 3 E). We explored both read coverage and read-pair sanity across these 29 contigs using Tablet (Milne et al., 2010) to validate the co-assembly of nematode and *Wolbachia*-like segments, as de Bruijn graph assemblers can create chimaeric contigs. We found the contigs to be valid, contiguous regions of the genome. Even in cases such as scaffold00357 (Figure 3 E) where the nuclear and *Wolbachia* components had distinct read coverages, manual inspection of the presumed *Wolbachia*-nuclear junctions revealed no issues of inconsistent read pairing or inferred insert length. Segments with much higher coverage than the nuclear genome may be derived from collapse of dispersed repeat copies of the *Wolbachia* fragment. From these analyses we conclude that the *Wolbachia*-like fragments are not from an unsuspected live *Wolbachia* infection of *D. viviparus*, but are rather neutrally-evolving insertions of *Wolbachia* genome fragments into the nematode nuclear genome, and are relics of an ancient symbiosis, now lost. We have called the fragmented *Wolbachia* wDv, though, obviously, we have no evidence of an extant wDv organism (and in fact regard it as being extinct).

As a preliminary assessment of the phylogenetic age of the insertions, we screened an independent *D. viviparus* isolate for presence of *Wolbachia* gene fragments. We performed directed PCR and Sanger sequencing of *Wolbachia* gene fragments from a *D. viviparus* isolate maintained at the Moredun Institute, Edinburgh, isolated in Scotland in 2005. Both *ftsZ* and 16S rRNA fragments were amplified from this strain, and, when sequenced, were closely similar to the whole genome assembly-derived fragments, but differed by several substitutions (Figure 4 A, B). Comparison of the nuclear small subunit ribosomal RNA sequence from the assembly to those from *Dictyocaulus* species affirmed the species identification (Figure 4 C). We also screened the previous *D. viviparus* transcriptome assembly (Cantacessi et al., 2011) for *Wolbachia*-like fragments and identified six transcribed fragments (Table 6) that were derived from wDv, confirming presence of symbiont gene fragments in a third isolate.

These transcribed *Wolbachia*-like fragments might offer evidence for functional integration of the remnants of the wDv genome into the nuclear genome. We thus investigated each fragment for possible function. In four of five fragments deriving from protein-coding genes there were frameshifts and in-frame stop codons. None of the transcribed fragments showed evidence of splicing. One transcript, where the *Wolbachia*-like sequence was in the likely 3' UTR of a nematode gene (a homologue of *C. elegans* FRM-1), showed standard spliceosomal introns in the nematode-gene-like part, but the *Wolbachia* fragment itself was not spliced. Four of the transcript fragments were very short (500-600 bases, approximately one 454 read length).

**Relationships of the *Wolbachia* of *D. viviparus* to other *Wolbachia***

To identify the relationships of wDv, sequences from the *Wolbachia*-like contigs were added to a five-gene supermatrix (including 16S rDNA, *groEL*, *ftsZ*, *dnaA* and *coxA* loci) used previously for phylogenetic analyses of *Wolbachia* (Lefoulon et al., 2012). This matrix does not include data from all 14 recognised *Wolbachia* clades, as sequencing in most has been limited. wDv fragments corresponding to these genes were identified using BLAST and aligned with MUSCLE. We were not able to identify a *dnaA* gene in the *D. viviparus* assembly. We added to the alignment data from *w*Oo and available sequences from the *Wolbachia* from *Radopholus similis* (*w*Rs). Both RAxML, MrBayes and PhyloBayes analyses suggested that *w*Dv belongs to clade F, with strong branch support (Figure 5). The long terminal branch of *w*Dv compared to other *Wolbachia* in the same clade is likely to be a consequence of the accumulation of mutations in the wDv regions due to their insertion and subsequent neutral evolution in the nematode genome. *w*Oo was placed robustly within clade C as expected. Placement of *w*Rs was less definite as it clustered as a sister taxon to clade D, but on a long branch with low support. We were unable to recover the published phylogeny (Haegeman et al., 2009) with *w*Rs arising basally to other *Wolbachia*, even when the matrix was analysed with *w*Dv excluded (data not shown), and thus this previous finding may be a methodological artifact.

One genomic feature that distinguishes clade C and D *Wolbachia* is the absence of WO phage. WO phage are active temperate bacteriophage that are present in the sequenced clade A and B genomes, and that may mediate genetic transfer of key symbiosis genes between strains (Kent and Bordenstein, 2010). Using the 1363 protein sequences derived from WO phage available in the NCBI/ENA/DDBJ databases we identified 15 scaffolds in the *D. viviparus* genome that contained significant (BLAST E-values less than 1e-20) to WO phage proteins. These matches (Table 7) were to a wide range of WO phage genes, including capsid proteins, portal proteins, secretion system components, recombinases and others. In this genomic feature, wDv resembled A and B *Wolbachia* more than it did C and D.

*Discussion*

**Fossils of *Wolbachia* infection reveal an unexpected palaeosymbiosis**

*D. viviparus* is the first nematode from the Rhabditina (the group that includes *C. elegans* and the important animal-parasitic Strongyloidea) that has been shown to have a relationship with *Wolbachia*. However, the *Wolbachia* sequences identified in the draft genome sequence do not appear to derive from a living organism, but rather show features suggestive of being ancient laterally transferred fragments of the genome of a clade F-like *Wolbachia*, which is now extinct. The insertions were not unique to the individual Cameroon nematode sampled, but were identified in another *D. viviparus* (from Scotland). Published and unpublished transcriptome data for *D. viviparus* include a very low level of fragments that mapped to *Wolbachia*-like regions of our assembly. We suggest that the lateral transfers may be found in all *D. viviparus*, and that it will be exciting to survey additional Dictyocaulinae and related families within Strongyloidea for evidence of (palaeo-) symbiosis, and to better date the origin of the laterally-transferred fragments.

Lateral transfers of *Wolbachia* DNA, *wont*s, have been identified previously in filarial nematodes and arthropods (Dunning Hotopp et al., 2007). The evidence for the *D. viviparus Wolbachia*-like sequences being ancient lateral transfers include their fragmentation, their interspersion with nematode sequence in robustly-assembled contigs, and their having inactivating mutations suggestive of neutral evolution of non-functional sequence. Read coverage of the *Wolbachia*-like fragments varied greatly. If all the fragments derived from the genome of a live infection, it would be expected that they would have very similar coverage, as seen in other *Wolbachia* infected nematodes (Godel et al., 2012, Kumar et al., 2013). While about 1 Mb of contigs had matches to *Wolbachia*, these did not constitute a complete genome. Only ~60% of the expected *Wolbachia* gene content was present (for example the *dnaA* gene was missing) and many genes and gene fragments were duplicated. Fragments with very high read coverage may be repeated (within the nematode genome), again suggestive of lack of live function. Genome fragmentation and neutral inactivation is suggestive of a long period of residence in the *D. viviparus* nuclear genome. We found six transcript fragments that included *Wolbachia* sequence, but closer analysis of these did not immediately suggest that they derived from functional genes, as they contained inactivating mutations or were incomplete fragments.

In the absence of a live *Wolbachia* organism, do these *Wolbachia*-like genes have a current expressed function in *D. viviparus*? The majority of the potential protein-coding genes in the *Wolbachia*-like fragments contain insertions, deletions, frameshift mutations or nonsense codons compared to their homologues from living *Wolbachia* genomes. We identified only six *Wolbachia*-like transcript fragments in 61,134 transcripts assembled from 3 million *D. viviparus* transcriptome sequences (Cantacessi et al., 2011). Four of the transcript fragments were very short, about one 454 read length, and one *Wolbachia* match was in the 3' untranslated region of a *bona fide* nematode gene. Four of five fragments from protein-coding genes had frameshift and in frame stop codon mutations, while the 16S rRNA fragment had a large deletion compared to 16S from living *Wolbachia*. As fragmented or degraded genes can still be active and expressed (since the function of a gene is not necessarily related to its full length), this analysis cannot be conclusive concerning gene function. We note that the *D. viviparus* RNA-Seq data were generated from poly(A)-selected RNA, and thus would not be expected to capture bacterial transcripts. Further studies will determine whether these and other *Wolbachia*-derived genes play roles in *D. viviparus* biology.

This discovery suggests that all three suborders of the nematode order Rhabditida (Rhabditina, Tylenchina and Spirurina) have members whose genomes and biology have been shaped by symbioses with *Wolbachia*. In the well-studied clade C and D *Wolbachia* the relationship has features of mutualism (Fenn and Blaxter, 2004). The *Wolbachia* observed in *R. similis* is apparently live, as bacterial cells can be seen within host cells by microscopy (Haegeman et al., 2009), but there are currently no data on the nature of the symbiosis: its genome sequence is awaited with interest. In *D. viviparus* we have no positive evidence for live infection. Our analyses placed both *w*Dv and *w*Rs close to clade F *Wolbachia*, and showed that clades C, D and F form a group distinct from clades A and B. Clade F appears more "promiscuous" in its host relationships (its known hosts include both nematodes and arthropods). In *Cimex*, the clade F symbiont may be essential for fertility and nymphal development. Clade F *Wolbachia* thus emerge as a credible source of the clade C and D *Wolbachia* of filarial nematode species. The wDv genome was likely to have contained WO phage (Kent and Bordenstein, 2010), a mobile element present in clade A and B genomes but strikingly absent from clade C and D genomes.

In this scenario, the genomic fossils of *Wolbachia* found in *D. viviparus* are the echo of a hit-and-run infection of an F-like *Wolbachia* in a dictyocauline ancestor. We note that there are *Wolbachia*-like sequences in transcriptome data from *A. caninum*, another strongyloidean nematode, and thus it is possible that *Wolbachia* infections may have been widespread in this group. While reports of *Wolbachia* in the strongyloidean *Angiostrongylus* have been discounted (Tsai et al., 2007, Foster et al., 2008), we are excited by the possibility that other palaeosymbioses, now extinct, may be revealed in forthcoming genome projects across the Nematoda and Metazoa.

Finally, we provide a first draft assembly and annotation of the important nematode parasite *D. viviparus*. The identification of our specimen as *D. viviparus* is based on close identity of sequenced loci and the complete mitochondrial genome between our specimen and previously published *D.*

*viviparus* data. As the specimen was destroyed during DNA extraction we no longer have a voucher for the individual. We note that there are very few records of *D. viviparus* in sub-Saharan Africa, and it is typically described as a temperate species (Thamsborg et al., 1998). A very large abattoir survey in the Democratic Republic of Congo found only 3 infected carcasses from 571 examined, and all of these were from cattle reared above 1,500 m (Ngaoundéré is at 1,200 m) (Chartier, 1990). The genome and annotation can be used as a springboard for further analysis both investigating the *Wolbachia*-nematode interaction and also potential gene identification for drug and vaccine development.

*Materials and methods*

## Nematode isolation and genome sequencing

A single *Dictyocaulus viviparus* male was isolated from *Bos indicus* (an individual of the local Gudali breed) in Ngaoundéré abattoir, Adamawa Region in Cameroon by David Ekale and Vincent Tanya during the ongoing Enhancing Protective Immunity Against Filariasis EU-Africa programme. The nematode was frozen at -80 °C and shipped to Liverpool, UK, where DNA was extracted using the DNeasy Blood & Tissue Kit (Qiagen). Genomic sequencing was carried out by the Edinburgh Genomics Facility, using Illumina TruSeq library preparation reagents and a HiSeq 2500 instrument. A single 300 bp insert library was constructed, and 100 base paired-end data generated. Raw data have been submitted to the International Nucleotide Sequence Database Consortium under the project accession PRJEB5116 (study ERP004482).

## Genome assembly and annotation

All software tools used (including versioning and command line options used) are summarised in Table 3. The quality of Illumina reads was checked with FASTQC (Andrews, 2013). Raw reads were quality trimmed (base quality of 20), and paired reads were discarded if either pair was below 51 bases using fastq-mcf (Aronesty, 2011). The trimmed reads were digitally normalised to ~20X coverage with khmer (Brown et al., 2012). A draft assembly was generated using the normalised reads with Velvet (Zerbino and Birney, 2008) and gaps within scaffolds were filled using GapFiller (Nadalin et al., 2012). Scaffold coverage was obtained by mapping all the reads back to the assembly using the clc-bio toolkit (CLC-Bio Ltd). Taxon-annotated GC%-coverage plots (TAGC plots) (Kumar et al., 2013) were used to identify potential bovine and other contamination. Bovine contamination, which was minimal, was removed.

A MAKER2-Augustus annotation pipeline was used to predict protein-coding genes from the genome (Cantarel et al., 2008). The MAKER2 program combines multiple *ab initio* and evidence-based gene predictors and predicts the most likely gene model. MAKER2 was run in a SGE cluster using the SNAP *ab initio* gene finder trained by CEGMA (Parra et al., 2007) output models, GeneMark-ES *ab initio* finder, *D. viviparus* transcripts and SwissProt proteins. We used the MAKER2 predictions to train Augustus (Stanke et al., 2006) and create a gene finder profile for *D. viviparus*. Using the gene finder profile, the assembled transcriptome (Cantacessi et al., 2011) and available expressed sequence tag data (Ranganathan et al., 2007), Augustus was used alone to predict the final gene set, which was used for downstream analysis. Protein sets from selected nematode species, downloaded from Wormbase (Harris et al., 2013), were clustered using orthoMCL (Li et al., 2003).

**Analysis of *Wolbachia*-like fragments**

The *Dictyaulus viviparus* draft assembly was broken into 500 bp fragments and each fragment was compared to *Brugia malayi* and *Onchocerca ochengi Wolbachia* endosymbiont genomes using BLAST+ (Boratyn et al., 2013). Similarity hits with lengths above 100 bases were considered for downstream analysis. Contigs with *Wolbachia*-like sequences were annotated using the RAST server, which provided both gene finding and gene functional annotation. Junction fragments between putative *Wolbachia* insertions and *D. viviparus* nuclear genomic DNA were identified using BLAST+. Putative phage WO fragments were identified through tBLASTn comparison of the 1353 phage WO proteins available in NCBI nr to the *D. viviparus* assembly, using an E-value cutoff of 1e-20.

The phylogenetic relationships of *Wolbachia* from *D. viviparus* were assessed by identifying orthologues of 16S rDNA, *groEL*, *ftsZ*, *dnaA*, and *coxA* genes, and aligning these to orthologues from other *Wolbachia*. The five-gene supermatrix was analysed using RAxML (Stamatakis, 2006), MrBayes (Ronquist et al., 2012) and PhyloBayes (Lartillot et al., 2009) (see Table 3 for specific parameters used). Trees were visualised in iTol (Letunic and Bork, 2011) and FigTree (Rambaut, 2012).

**Identification of *Wolbachia* insertions in other *D. viviparus***

*D. viviparus* genomic DNA from the Moredun, Scotland, isolate was provided by Prof. Jacqui matthews, Moredun Institute (Pezzementi et al., 2012). The Moredun strain has no known connection with Cameroon. *Caenorhabditis elegans* (free-living rhabditid nematode, which does not carry *Wolbachia*) and *Litomosoides sigmodontis* (a filarial nematode that carries a clade D *Wolbachia* (Hoerauf et al., 1999)) genomic DNAs were used as negative and positive controls, respectively. PCR primers designed to amplify *Wolbachia* 16S, *Wolbachia ftsZ* (Bandi et al., 1998), nematode nuclear small subunit rRNA (nSSU) (Blaxter et al., 1998) and mitochondrial cytochrome oxidase I (*cox1*) (Folmer et al., 1994) were used in PCR with Phusion enzyme (NEB) to identify similar fragments in each nematode genomic DNA. A list of primers used and PCR conditions are given in Table 4. Positive PCR fragments were directly sequenced in both directions using BigDye v3 reagents in the Edinburgh Genomics facility. *D. viviparus* Roche 454 transcriptome data (Bioproject PRJNA20439) were downloaded from ENA and screened using BLAST for sequences corresponding to the *Wolbachia* insertions in our assembly.

*Acknowledgements*

We thank David Ekale for collection of the material in Cameroon, Catherine Hartley in Liverpool for the DNA extraction, Jacqui Matthews for Moredun *D. viviparus* DNA, Simon Babayan for *L. sigmodontis* DNA, and Karim Gharbi and Anna Montazam of the Edinburgh Genomics Facility for expert library preparation and sequencing. GK is supported by the BBSRC through a PhD studentship, and the School of Biological Sciences, University of Edinburgh. This work was supported in part by the EU EPIAF programme (contract HEALTH-F3-2010-242131).

*Figure legends*

## Figure 1 Relationships of nematode species harbouring *Wolbachia* symbionts

A phylogenetic cartoon showing the relationships of the nematode species discussed in this work (Blaxter, 2011). To the left, the systematic structure of the class Chromadoria is given, and the three major suborders within Rhabditida are highlighted. Lifecycle strategies of the groups are indicated. The fine-scale relationships of species discussed in the text are given to the right. The presence of live *Wolbachia* infection (+ : yes, - : no), evidence of laterally-transferred *Wolbachia* sequences in the nuclear genome (+ : yes, - : no, ? : unknown), and the availability of complete genome sequences (+ : yes, - : no, ± : partial genome sequence) for each of the species are indicated.

## Figure 2 Comparison of the *Dictyocaulus viviparus* proteome to that of other rhabditid nematodes

Venn diagram illustrating the orthoMCL clustering of the predicted proteome of *Dictyocaulus viviparus* (DVI) to those of *Caenorhabditis elegans* (CEL), *Heterorhabditis bacteriophora* (HBA) and *Haemonchus contortus* (HCO). The numbers of proteins clustered and the total number of predicted proteins is given below each species' name.

## Figure 3 *Wolbachia* sequence in a *Dictyocaulus viviparus* genome assembly

**A** Taxon-annotated GC%-coverage plot of the primary *D. viviparus* genome assembly, with contigs that have significant matches to *Wolbachia* proteins highlighted in red. A total of 193 contigs spanning 1 Mb (out of a total assembly span of 169 Mb) had significant similarity to *Wolbachia*.

**B** Circos plot comparing the 25 longest of the *D. viviparus* genome contigs that contained *Wolbachia*-like sequence to the genome of the *Wolbachia* endosymbionts of the filarial nematode *Brugia malayi* (*w*Bm) (Foster et al., 2005) and *Onchocerca ochengi* (*w*Oo). The arcs show BLASTn-derived matches between the contigs and the genome sequences. Transcripts from *D. viviparus* mapped to the assembly are reported as green lines in the outer circle of the figure.

**C** Frequency histogram illustrating the different patterns of coverage of the *Wolbachia*-like scaffolds (black) compared to the nuclear genome scaffolds (green).

**D** Frequency plot of similarity of *D. viviparus Wolbachia*-like sequences to *w*Bm (blue) and *w*Oo (the *Wolbachia* endosymbiont of the filarial nematode *Onchocerca ochengi*) (red). Each *D. viviparus*

*Wolbachia*-like segment was split into 500 bp fragments, and the best percentage identity with the reference genomes calculated using BLASTn.

**E** The *Wolbachia*-like fragments identified in the *D. viviparus* genome assembly are co-assembled with nematode genes, and have accumulated multiple inactivating mutations. Two putative *Wolbachia* insertions in nuclear contigs are shown in views derived from the gBrowse genome viewer. Each panel shows (from top to bottom) the whole scaffold with the zoomed-in region highlighted, the GC% plot for the scaffold, the scale for the zoomed-in region, the read coverage for the zoomed-in region, the genes called by RAST in the zoomed in region and the genes called by AUGUSTUS in the zoomed -in region. The upper plot shows scaffold00357 while the lower plot shows scaffold00506.

**Figure 4 Comparison of *Wolbachia*-like insertions from two *Dictyocaulus viviparus* isolates, and relationships of the Cameroon *D. viviparus***

**A** 16S rRNA gene fragments from the Cameroon isolate of *D. viviparus* (obtained through whole genome sequencing) and from the Moredun isolate (from specific amplification) are shown aligned. The genome sequence assembly has three copies of *Wolbachia*-like 16S genes, two tandemly arranged and truncated in scaffold scaf09320, and one in scaffold scaf01523.

**B** *ftsZ* gene fragments from the Cameroon isolate of *D. viviparus* (obtained through whole genome sequencing) and from the Moredun isolate (from specific amplification) are shown aligned. While we were able to amplify the complete fragment from the Moredun strain, the genome assembly contains only a truncated *ftsZ* gene (and no consensus is shown for the ~200 bases of essentially unaligned sequence at the 5' end of the alignment).

**C** Bayesian phylogenetic analysis of the complete nuclear small subunit ribosomal RNA (nSSU) genes of the Cameroon *D. viviparus* and other *Dictyocaulus* sp., and outgroups (taken from the European Nucleotide Archive). The Cameroon *D. viviparus* is most similar to the European *D. viviparus* sequenced previously. RAxML analyses yielded the same topology. The 5' gene fragment isolated and sequenced from the Moredun strain was identical to the other *D. viviparus* nSSU sequences.

**Figure 5 Analysis of the phylogenetic relationships of the *Wolbachia* nuclear insertions in the *Dictyocaulus viviparus* genome**

Phylogenetic tree inferred from 16S rDNA, *groEL, ftsZ, dnaA* and *coxA* loci with maximum likelihood (RAxML) and Bayesian (MrBayes, PhyloBayes) inference. Branch support is reported as (RaxML/MrBayes/PhyloBayes). Strains representing *Wolbachia* supergroups A, B, C, D, F and H are indicated.

*Tables*

**Table 1 Assembly statistics for the *Dictyocaulus viviparus* nuclear genome and the *Wolbachia*-like insertions.**

|  | *D. viviparus* nuclear genome * | *Wolbachia*-like fragments ** |
|---|---|---|
| **number of reads (million)** | 165 |  |
| **span of data (Gb)** | 16 |  |
| **span of assembly (Mb)** | 169.4 | 1.0 |
| **number of contigs** | 17,715 | 193 |
| **N50 length (bp)** | 22,560 | 10,017 |
| **mean read coverage** | 84.53 | 119.06 |
| **GC%** | 34.5 | 34.9 |

\* The *D. viviparus* mitochondrial genome was assembled in four contigs, with mean coverage ~10,000 fold. The four contigs were aligned to the published *D. viviparus* mitochondrion genome and cover the entire span.

\*\* Fragment lengths were added as full contigs if no nematode-like sequence was detected. If the contig contained nematode sequences, only the range of the *Wolbachia* BLAST hits was added.

**Table 2 Genome assembly and annotation metrics of *D. viviparus* and other Rhabditina species**

| Species | *Dictyocaulus viviparus* | *Caenorhabditis elegans* | *Haemonchus contortus* | *Heterorhabditis bacteriophora* |
|---|---|---|---|---|
| **Assembly size (Mb)** | 169.4 | 100.3 | 368.8 | 77.0 |
| **Number of contigs >500 bp** | 17715 | 6 | 19728 | 1259 |
| **Mean contig length >500 bp** | 9561 | 14326628 | 18696 | 61164 |
| **N50 >500 bp** | 22560 | 17493829 | 83501 | 312328 |
| **GC** | 34.5 | 35.4 | 43.1 | 33.3 |
| **Number of N's (Mb)** | 0.5 | 0 | 23.6 | 2.6 |
| **Predicted genes** | 14306 | 20520 | 21276 | 14667 |
| **Median protein length (bp)** | 834 | 1017 | 900 | 423 |
| **Median exon length (bp)** | 168 | 146 | 109 | 94 |
| **Median exons per gene** | 7 | 5 | 7 | 4 |
| **Reference** | this work | (The *C. elegans* Genome Sequencing Consortium, 1998) | (Schwarz et al., 2013, Laing et al., 2013) | (Bai et al., 2013) |

**Table 3 Putative *Wolbachia*-like open reading frames identified in the *Dictyocaulus viviparus* nuclear genome**

| Feature | Value | Comment |
|---|---|---|
| Number of open reading frames (ORFs) * | 1580 | |
| Mean ORF length | 729 ± 703 bp | In *w*Bm the mean length is 859 ± 712 bp |
| Distinct *Wolbachia* genes identified ** | 567 | These are present in 1033 ORFs. 547 ORFs had no similarity to other *Wolbachia* genes. |
| Genes identified in only 1 ORF | 318 | 134 had < 70% coverage; 79 of these genes are not present in *w*Bm |
| Genes identified in more than 1 ORF | 249 | Mean number of ORFs per gene identifier = 2.9; SD = 1.4 |

* Predicted using RAST. The RAST analysis of the *Wolbachia*-like fragments from *D. viviparus* is available on the RAST server at http://rast.nmpdr.org/?page=JobDetails&job=112231.

** These are genes identified by RAST as being similar to genes identified in other *Wolbachia* genomes. Some genes are present in multiple, distinct copies in the *D. viviparus* assembly.

**Table 4 Analysis software versions and parameter settings**

| Software tool | Reference | Version | Parameters used* | Comments |
|---|---|---|---|---|
| FASTQC | (Andrews, 2013) | v0.10.1 | | |
| fastq-mcf | (Aronesty, 2011) | ea-utils.1.1.2-537 | -l 51 -q 20 --qual-mean 20 -R | |
| Blobology | (Kumar et al., 2013) | 2013-10-21 | default | |
| Khmer | (Brown et al., 2012) | khmer-17-05-2013 | -k 20 -C 20 -p | |
| Velvet | (Zerbino and Birney, 2008) | 1.2.08 | -exp_cov auto -cov_cutoff auto | Kmer length of 51 was used |
| GapFiller | (Nadalin et al., 2012) | v1-11 | -o 10 -m 55 | |
| clc_bio | program used: clc_mapper | 4.1.0 | -l 0.9 -s 0.9 | |
| BLAST | (Boratyn et al., 2013) | 2.25 | default | |
| CEGMA | (Parra et al., 2007) | 2.0 | default | |
| SNAP | (Korf, 2004) | 2006-07-28 | default | used within MAKER pipeline |
| GeneMark | (Lomsadze et al., 2005) | v.2.3e | --BP OFF -max_nnn 500 -min_contig 10000 | |
| MAKER2 | (Cantarel et al., 2008) | 2.25 | default | maker_opts file changed |
| Augustus | (Stanke et al., 2006) | 2.7 | script used: | |

| | | | auto_Aug.pl | |
|---|---|---|---|---|
| orthoMCL | (Li et al., 2003) | 2.0.3 | default | |
| MUSCLE | (Edgar, 2004) | 3.8.31 | default | |
| RAxML | (Stamatakis, 2006) | 7.6.4 | -m GTRGAMMA | |
| MrBayes | (Ronquist et al., 2012) | 3.2 | lset nst= 6 rates= gamma | |
| PhyloBayes | (Lartillot et al., 2009) | 2.3 | -cat -gtr | |
| FigTree | (Rambaut, 2012) | 3.0.2 | | used in construction of Figure 3 C and Figure 4 |
| iTOL | (Letunic and Bork, 2011) | | | used in construction of Figure 3 C and Figure 4 |
| Geneious | www.geneious.com | R7 | | used for construction of Figure 3 A, B |

\* Unless otherwise specified, default parameters were used.

**Table 5 PCR test for *Wolbachia* insertions**

| Target gene | Primer F (name, sequence 5' to 3') | Primer R (name, sequence 5' to 3') | *Dictyocaulus viviparus*\* | *Caenorhabditis elegans*\* | *Litomosoides sigmodontis*\* | Reference for primers |
|---|---|---|---|---|---|---|
| *Wolbachia* 16S rRNA | Wspec16S_F1 GAAGATAATGACGGTACTCAC | Wspec16S_R1 GTCACTGATCCCACTTTAAATAAC | + | - | + | (Bandi et al., 1998) |
| *Wolbachia* ftsZ | ftsZ_F1 ATYATGGARCATATAAARGATAG | ftsZ_R1 TCRAGYAATGGATTRGATAT | + | - | + | (Bandi et al., 1998) |
| nuclear nSSU | F04 GCTTGTCTCAAAGATTAAGCC | R26 CATTCTTGGCAAATGCTTTCG | + | + | + | (Blaxter et al., 1998) |
| mitochondrial cox1 | LCO1490 GGTCAACAAATCATAAAGATATTGG | HCO2198 TAAACTTCAGGGTGACCAAAAAATCA | + | + | + | (Folmer et al., 1994) |

\* + strong positive band observed, and sequence confirmed; - no PCR product observed. All PCRs used New England BioLabs Phusion HF mix, an annealing temperature of 58 °C, 35 cycles of amplification, and were repeated twice with identical results.

**Table 6 Possible transcribed genes of *Wolbachia* origin in the *Dictyocaulus viviparus* genome**

| name of transcript fragment * | length (bp) | BLASTn E-value of best match to *Wolbachia* genomes | functional identification of matched gene | genomic scaffold(s) containing match(es) | frameshift mutations | in frame stop codons | comments |
|---|---|---|---|---|---|---|---|
| 18187 | 3523 | 5.00E-160 | inosine monophosphate dehydrogenase | nDv.1.0.scaf06859 nDv.1.0.scaf15124 | 2 | 0 | This long transcript fragment contains a nematode gene on one strand (a homologue of *C. elegans* FRM-1) and a short match to the N-terminus of *Wolbachia* IMP dehydrogenase on the other. The *D. viviparus* FRM-1 gene is on a scaffold (06859) that also contains the N-terminal *Wolbachia* IMP dehydrogenase fragment. Additional matches to IMP dehydrogenase (not in the transcript fragment) are found on scaffold 15124. |
| 31017 | 1146 | 0 | chaperonin GroEL | nDv.1.0.scaf17518 nDv.1.0.scaf05527 | 3 | 0 | There are two matches to *Wolbachia* GroEL in the nuclear assembly. The transcript fragment matches scaffold 17518 better (96% identity) than it does 05527, but 05527 contains a longer GroEL fragment. |
| 34819 | 660 | 1.00E-125 | 30S ribosomal protein S6 | nDv.1.0.scaf07136 | 0 | 0 | The transcript fragment appears to be a full-length copy of the S6 gene, without frameshifts. The transcript fragment matches a repetitive nuclear scaffold that contains two partial copies of an S6-like gene. |
| 35543 | 621 | 9.00E-146 | 50S ribosomal protein L25 | nDv.1.0.scaf04587 | 2 | 0 | The match in the nuclear assembly extends beyond the transcript fragment, and includes additional frameshifting indels. |
| 36721 | 569 | 9.00E-136 | phosphoglyceromutase | nDv.1.0.scaf00055 | 1 | 1 | The transcript fragment covers only the first 40% of the *Wolbachia* phosphoglyceromutase. The nuclear assembly includes a fuller length match with additional frameshifts and indels. |
| 38836 | 512 | 1.00E-33 | 16S rRNA | nDv.1.0.scaf01523 | | | A short fragment of the 16S RNA**. The match to scaffold 01523 is not perfect, as it includes many substitutions and an 18 base indel. |

\* The transcriptome assembly is from 454 data of (Cantacessi et al., 2011).
\*\* This fragment not from the same region that was amplified from the Moredun strain.

**Table 7 Matches to *Wolbachia* WO phage in the *Dictyocaulus viviparus* genome assembly**

| *Dictyocaulus viviparus* genome scaffold | Accession of matched *Wolbachia* WO phage protein* | Description of matched *Wolbachia* WO phage protein* | % identity | length of match | start point of match in D. vivaparus scaffold | end point of match in D. vivaparus scaffold | BLAST E-value | BLAST bitscore |
|---|---|---|---|---|---|---|---|---|
| **nDv.1.0.scaf00337** | ref:NP_965981 1 | phage SPO1 DNA polymerase-related protein [*Wolbachia* endosymbiont of *Drosophila melanogaster*] | 83.68 | 190 | 38869 | 39438 | 6.00E-107 | 333 |
| | gb:EEH12356 1 | phage portal protein HK97 family [*Wolbachia* endosymbiont of *Muscidifurax uniraptor*] | 45.67 | 254 | 23126 | 23836 | 3.00E-47 | 187 |
| | gb:EAL59902 1 | Phage portal protein [*Wolbachia* endosymbiont of *Drosophila simulans*] | 43.02 | 179 | 22954 | 23469 | 2.00E-39 | 116 |
| | ref:WP 006013219 1 | Phage portal protein HK97 family (fragment) [*Wolbachia pipientis*] | 58.33 | 60 | 22761 | 22940 | 1.00E-25 | 74.3 |
| **nDv.1.0.scaf00809** | dbj:BAH22317 1 | ankyrin motif protein [*Wolbachia* endosymbiont of *Cadra cautella*] | 26.57 | 271 | 18496 | 17795 | 1.00E-10 | 70.5 |
| **nDv.1.0.scaf01202** | ref:YP_007889020 1 | Phage contractile tail tube protein [*Wolbachia* endosymbiont of *Drosophila simulans* wHa] | 69.39 | 49 | 23532 | 23386 | 2.00E-13 | 68.2 |
| **nDv.1.0.scaf01523** | dbj:BAH22263 1 | replicative DNA helicase [*Wolbachia* endosymbiont of *Cadra cautella*] | 87.27 | 330 | 3105 | 4094 | 7.00E-154 | 543 |
| | | | 88.89 | 243 | 4090 | 4818 | 4.00E-180 | 405 |
| | gb:EAL60078 1 | phage host specificity protein [*Wolbachia* endosymbiont of *Drosophila simulans*] | 57.85 | 344 | 13385 | 14410 | 5.00E-96 | 352 |
| | | | 67.21 | 183 | 14350 | 14889 | 4.00E-149 | 262 |
| | | | 52.36 | 275 | 12200 | 12976 | 9.00E-54 | 211 |
| | | | 56.41 | 156 | 15041 | 15499 | 4.00E-149 | 155 |
| | | | 58.62 | 116 | 15538 | 15885 | 4.00E-149 | 121 |
| | | | 72.41 | 58 | 13201 | 13028 | 3.00E-14 | 80.9 |
| | | | 42.42 | 132 | 13228 | 13620 | 2.00E-13 | 77.8 |
| **nDv.1.0.scaf02083** | ref:NP_966244 1 | HK97 family phage major capsid protein [Wolbachia endosymbiont of Drosophila melanogaster] | 63.27 | 343 | 7439 | 6435 | 2.00E-117 | 418 |
| | emb:CCE77413 1 | putative phage major capsid protein HK97 family (part 2) [Wolbachia pipientis wAlbB] | 65.22 | 92 | 6710 | 6435 | 3.00E-31 | 123 |
| | ref:YP_002726909 1 | phage prohead protease [Wolbachia sp wRi] | 70.24 | 84 | 16117 | 15866 | 7.00E-35 | 114 |
| **nDv.1.0.scaf04054** | ref:YP_007889449 1 | Putative phage terminase [Wolbachia endosymbiont of Drosophila simulans wHa] | 63 | 100 | 5800 | 6093 | 3.00E-30 | 116 |
| | ref:YP_002727472 1 | phage uncharacterized protein [Wolbachia sp wRi] | 59.8 | 102 | 5549 | 5851 | 4.00E-24 | 112 |
| **nDv.1.0.scaf05447** | ref:WP 017532175 1 | phage portal protein [Wolbachia endosymbiont of Diaphorina citri] | 56.14 | 57 | 1907 | 2077 | 7.00E-11 | 64.7 |
| **nDv.1.0.scaf08784** | ref:YP_002727488 1 | site-specific recombinase phage integrase family [*Wolbachia* sp. wRi] | 71.64 | 134 | 943 | 542 | 4.00E-90 | 200 |
| | | | 70.09 | 107 | 1242 | 922 | 4.00E-90 | 152 |
| | gb:EEB56362 1 | site-specific recombinase phage integrase family [*Wolbachia* endosymbiont of *Culex quinquefasciatus* JHB] | 77.5 | 40 | 1384 | 1265 | 1.00E-09 | 63.5 |
| **nDv.1.0.scaf11970** | ref:NP_966849 1 | phage uncharacterized protein [*Wolbachia* endosymbiont of *Drosophila melanogaster*] | 65.31 | 147 | 1631 | 1191 | 1.00E-46 | 186 |
| **nDv.1.0.scaf12214** | dbj:BAH22270 1 | putative DNA recombinase [*Wolbachia* endosymbiont of *Cadra cautella*] | 43.17 | 139 | 141 | 533 | 2.00E-55 | 112 |
| | | | 35.98 | 189 | 529 | 1002 | 2.00E-55 | 103 |
| **nDv.1.0.scaf12590** | dbj:BAH22270 1 | putative DNA recombinase [*Wolbachia* endosymbiont of *Cadra cautella*] | 43.33 | 120 | 1188 | 838 | 1.00E-50 | 101 |
| | | | 33.86 | 189 | 842 | 369 | 1.00E-50 | 90.9 |

| nDv.1.0.scaf12848 | dbj:BAH22266 1 | putative type IV secretion system protein VirB8 [*Wolbachia* endosymbiont of *Cadra cautella*] | 60.78 | 204 | 344 | 955 | 7.00E-55 | 212 |
|---|---|---|---|---|---|---|---|---|
| | | | 57.14 | 70 | 1066 | 1275 | 2.00E-17 | 73.6 |
| | dbj:BAH22267 1 | hypothetical protein [*Wolbachia* endosymbiont of *Cadra cautella*] | 66.67 | 66 | 55 | 252 | 4.00E-19 | 94 |
| nDv.1.0.scaf13303 | dbj:BAH22266 1 | putative type IV secretion system protein VirB8 [*Wolbachia* endosymbiont of *Cadra cautella*] | 65.59 | 93 | 115 | 393 | 1.00E-20 | 98.6 |
| nDv.1.0.scaf13854 | ref:NP_965981 1 | phage SPO1 DNA polymerase-related protein [*Wolbachia* endosymbiont of *Drosophila melanogaster*] | 88.89 | 207 | 943 | 323 | 6.00E-108 | 389 |
| nDv.1.0.scaf14606 | ref:YP_007889449 1 | Putative phage terminase [*Wolbachia* endosymbiont of *Drosophila simulans* wHa] | 65.91 | 88 | 603 | 340 | 2.00E-28 | 126 |
| | | | 40.94 | 127 | 379 | 14 | 1.00E-14 | 80.5 |

\* Because the *Wolbachia* insertions in the *D. viviparus* genome were inactivated by mutation, many genes had multiple adjacent, independent high scoring segment matches in different frames in BLAST searches.

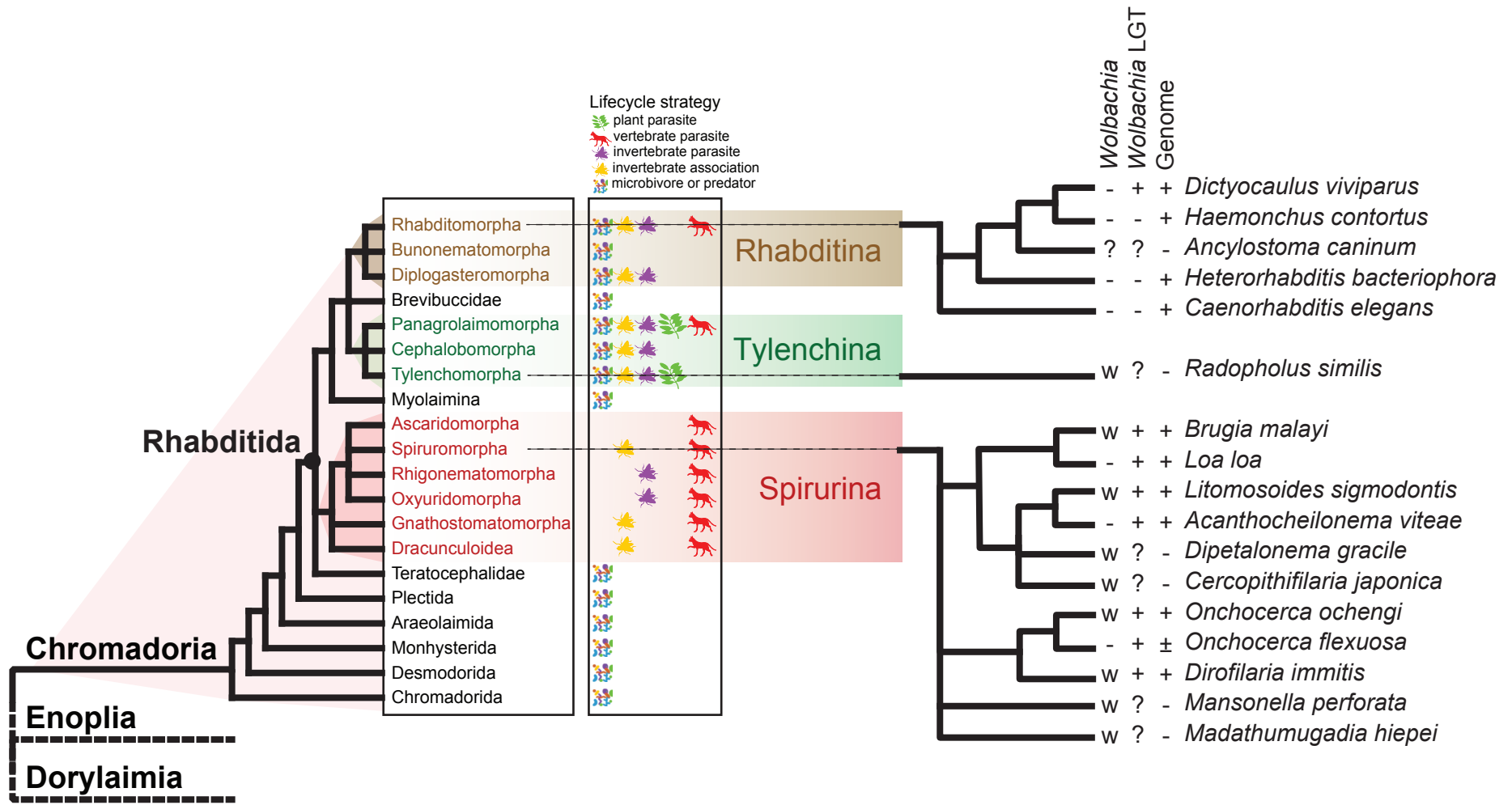

Figure 1 Relationships of nematode species harbouring Wolbachia symbionts
A phylogenetic cartoon showing the relationships of the nematode species discussed in this work (Blaxter, 2011). To the left, the systematic structure of the class Chromadoria is given, and the three major suborders within Rhabditida are highlighted. Lifecycle strategies of the groups are indicated. The fine-scale relationships of species discussed in the text are given to the right. The presence of live Wolbachia infection (+ : yes, - : no), evidence of laterally-transferred Wolbachia sequences in the nuclear genome (+ : yes, - : no, ? : unknown), and the availability of complete genome sequences (+ : yes, - : no, ± : partial genome sequence) for each of the species are indicated.

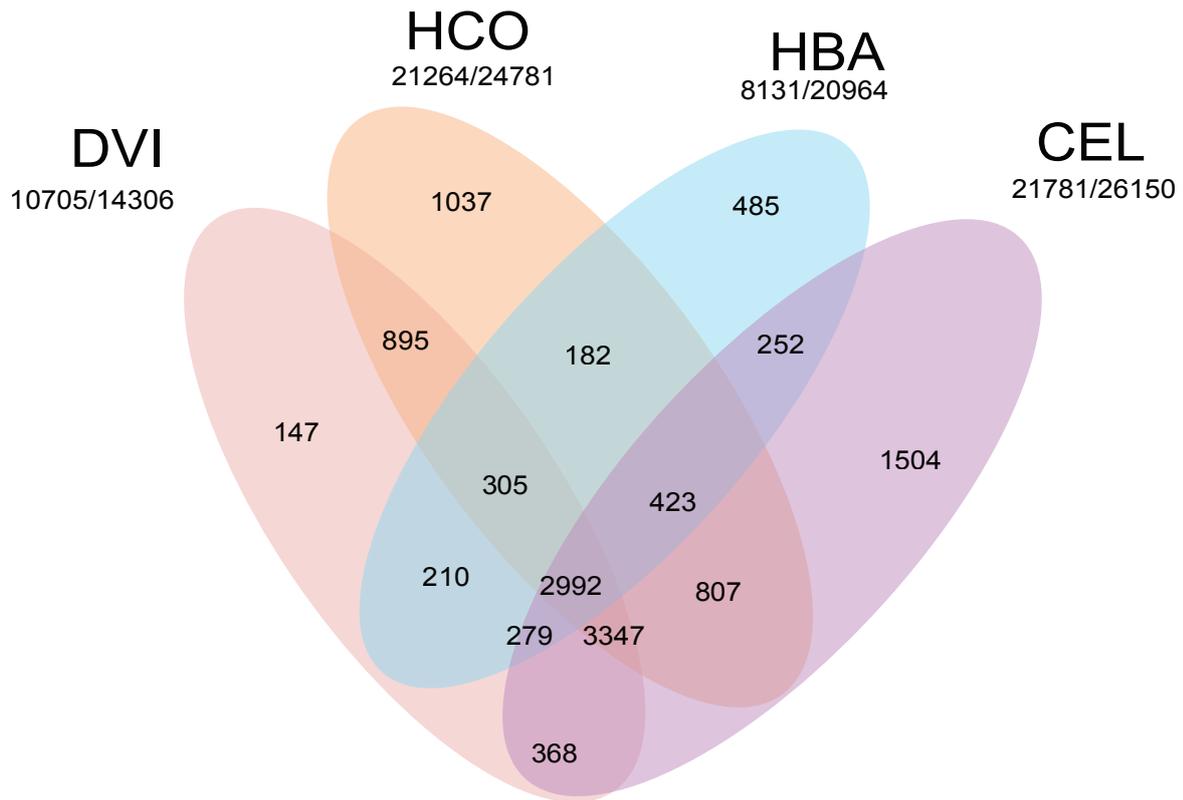

Figure 2 Comparison of the Dictyocaulus viviparus proteome to that of other rhabditid nematodes
Venn diagram illustrating the orthoMCL clustering of the predicted proteome of Dictyocaulus viviparus (DVI) to those of Caenorhabditis elegans (CEL), Heterorhabditis bacteriophora (HBA) and Haemonchus contortus (HCO). The numbers of proteins clustered and the total number of predicted proteins is given below each species' name.

**Figure 3** *Wolbachia* sequence in a *Dictyocaulus viviparus* genome assembly

**A** Taxon-annotated GC%-coverage plot of the primary *D. viviparus* genome assembly, with contigs that have significant matches to *Wolbachia* proteins highlighted in red. A total of 193 contigs spanning 1 Mb (out of a total assembly span of 169 Mb) had significant similarity to *Wolbachia*.

**B** Circos plot comparing the 25 longest of the *D. viviparus* genome contigs that contained *Wolbachia*-like sequence to the genome of the *Wolbachia* endosymbionts of the filarial nematode *Brugia malayi* (*w*Bm) (Foster et al., 2005) and *Onchocerca ochengi* (*w*Oo). The arcs show BLASTn-derived matches between the contigs and the genome sequences. Transcripts from *D. viviparus* mapped to the assembly are reported as green lines in the outer circle of the figure.

**C** Frequency histogram illustrating the different patterns of coverage of the *Wolbachia*-like scaffolds (black) compared to the nuclear genome scaffolds (green).

**D** Frequency plot of similarity of *D. viviparus Wolbachia*-like sequences to *w*Bm (blue) and *w*Oo (the *Wolbachia* endosymbiont of the filarial nematode *Onchocerca ochengi*) (red). Each *D. viviparus Wolbachia*-like segment was split into 500 bp fragments, and the best percentage identity with the reference genomes calculated using BLASTn.

**E** The *Wolbachia*-like fragments identified in the *D. viviparus* genome assembly are co-assembled with nematode genes, and have accumulated multiple inactivating mutations. Two putative *Wolbachia* insertions in nuclear contigs are shown in views derived from the gBrowse genome viewer. Each panel shows (from top to bottom) the whole scaffold with the zoomed-in region highlighted, the GC% plot for the scaffold, the scale for the zoomed-in region, the read coverage for the zoomed-in region, the genes called by RAST in the zoomed in region and the genes called by AUGUSTUS in the zoomed -in region. The upper plot shows scaffold00357 while the lower plot shows scaffold00506.

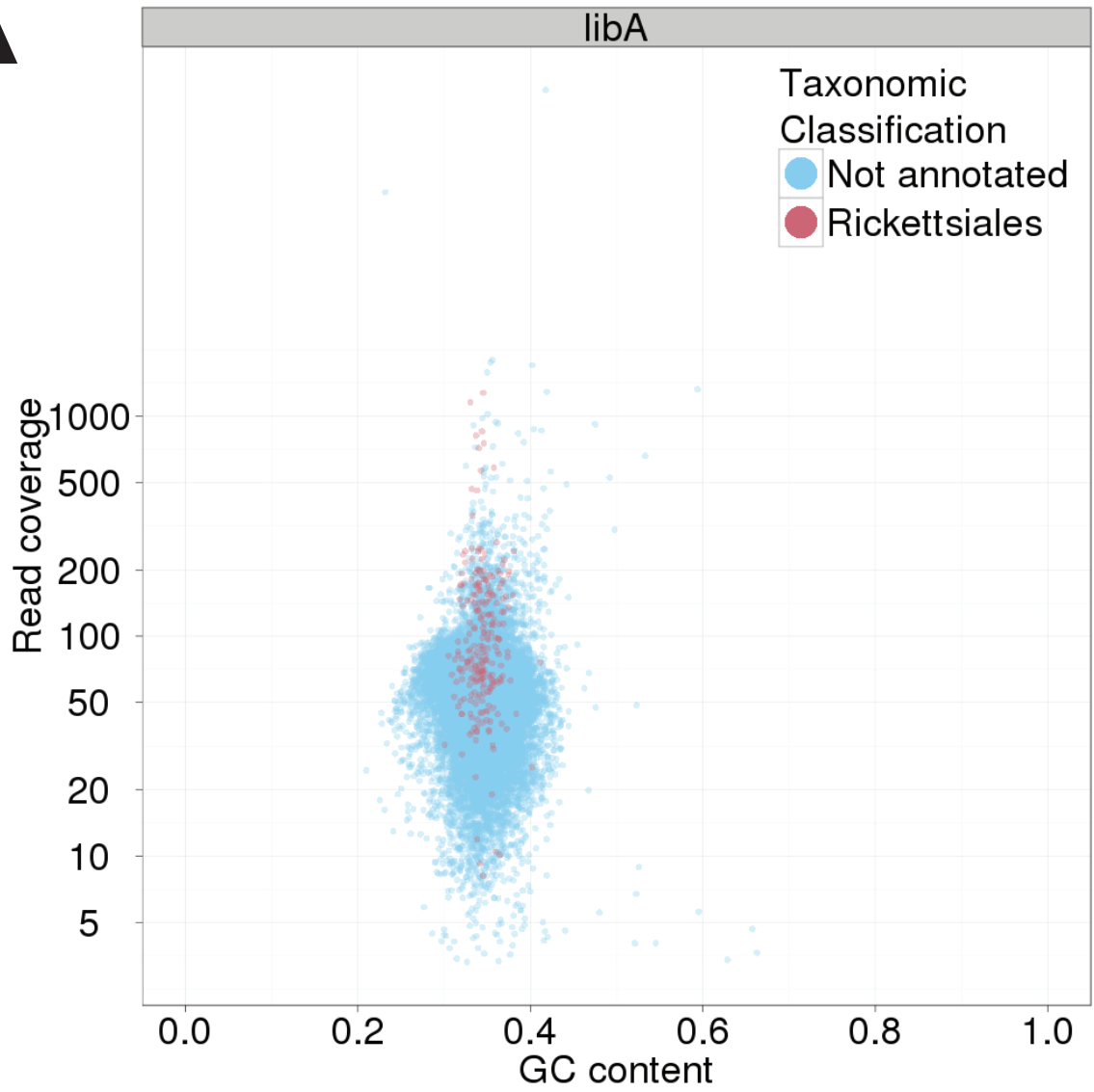

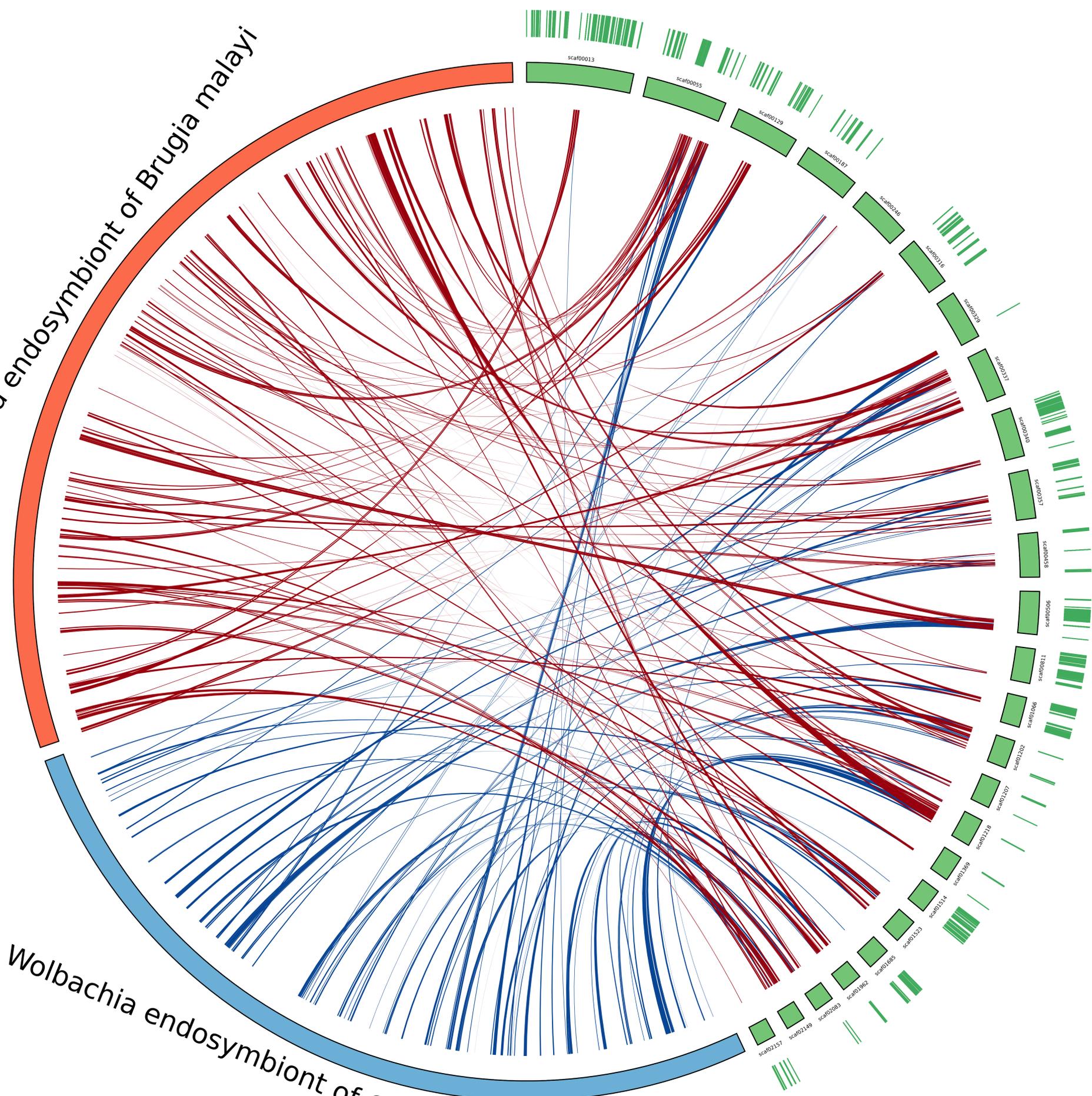

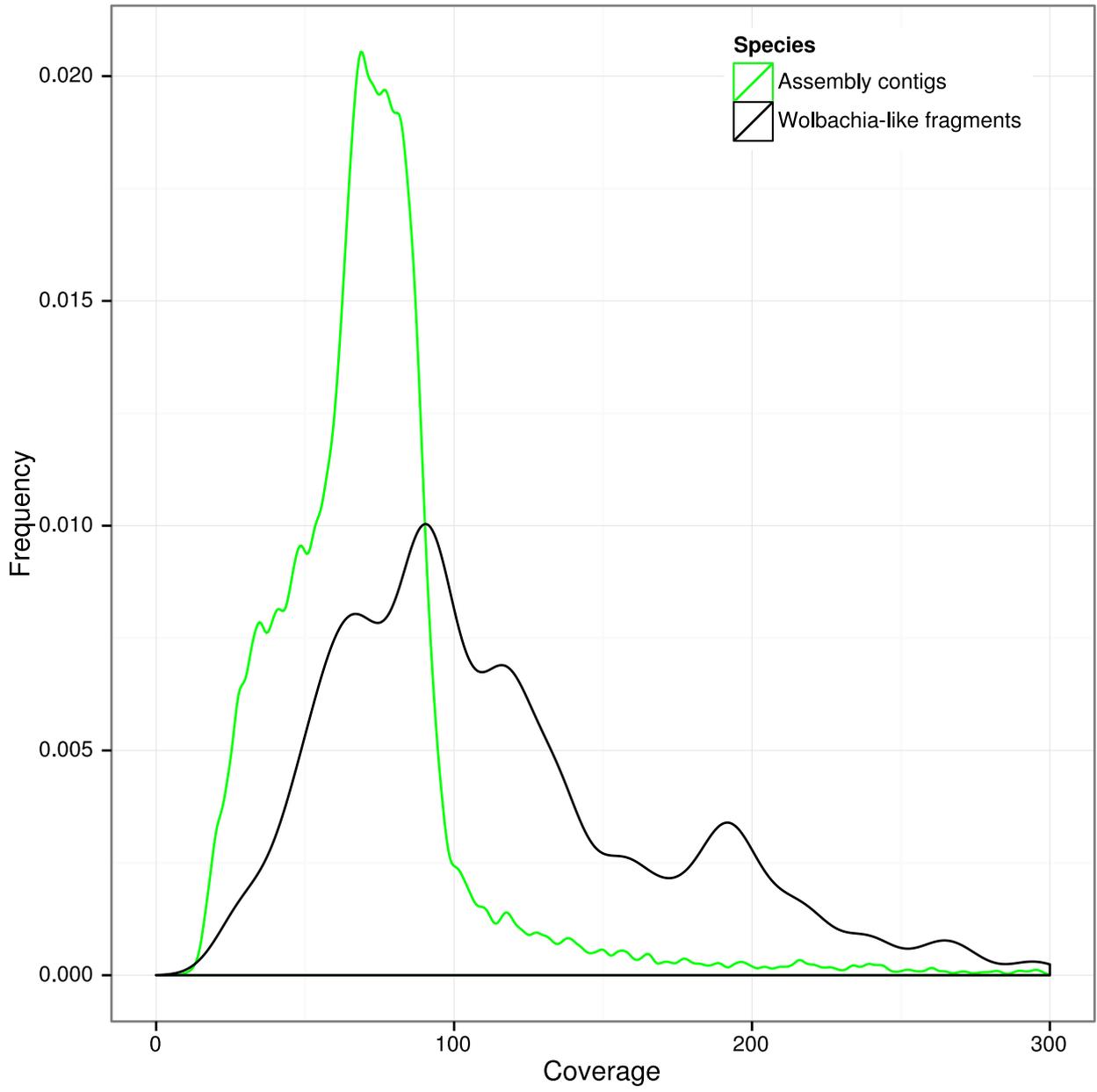

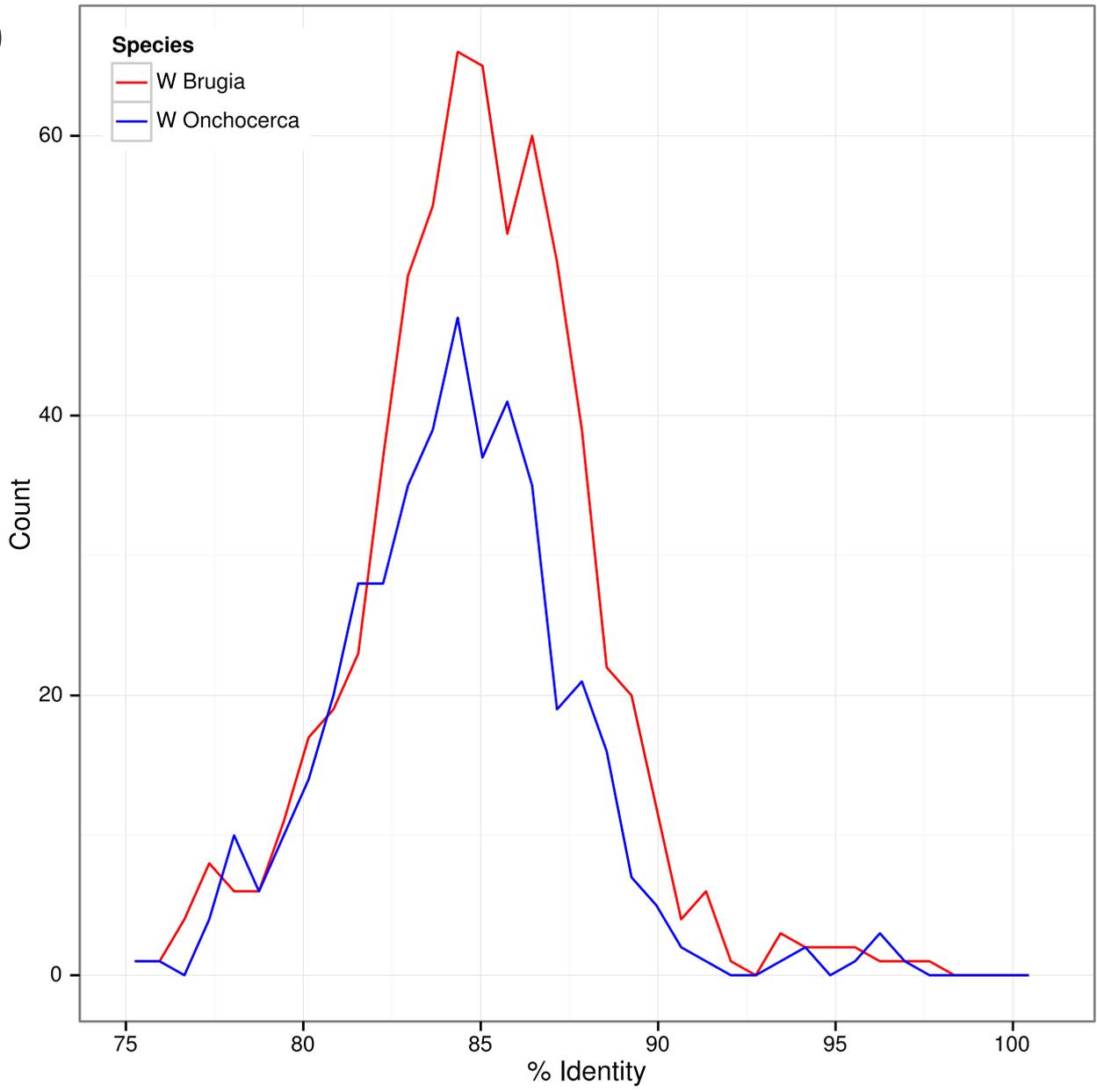

**E** ## scaffold00357

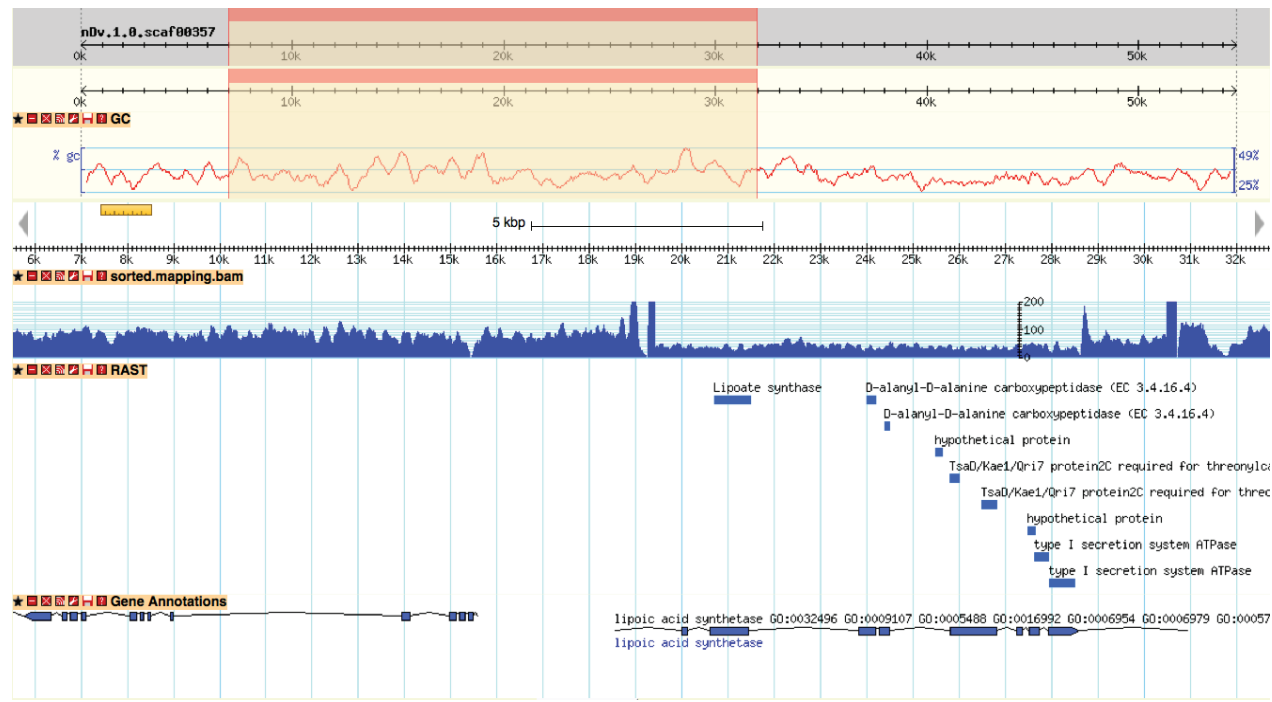

## scaffold00506

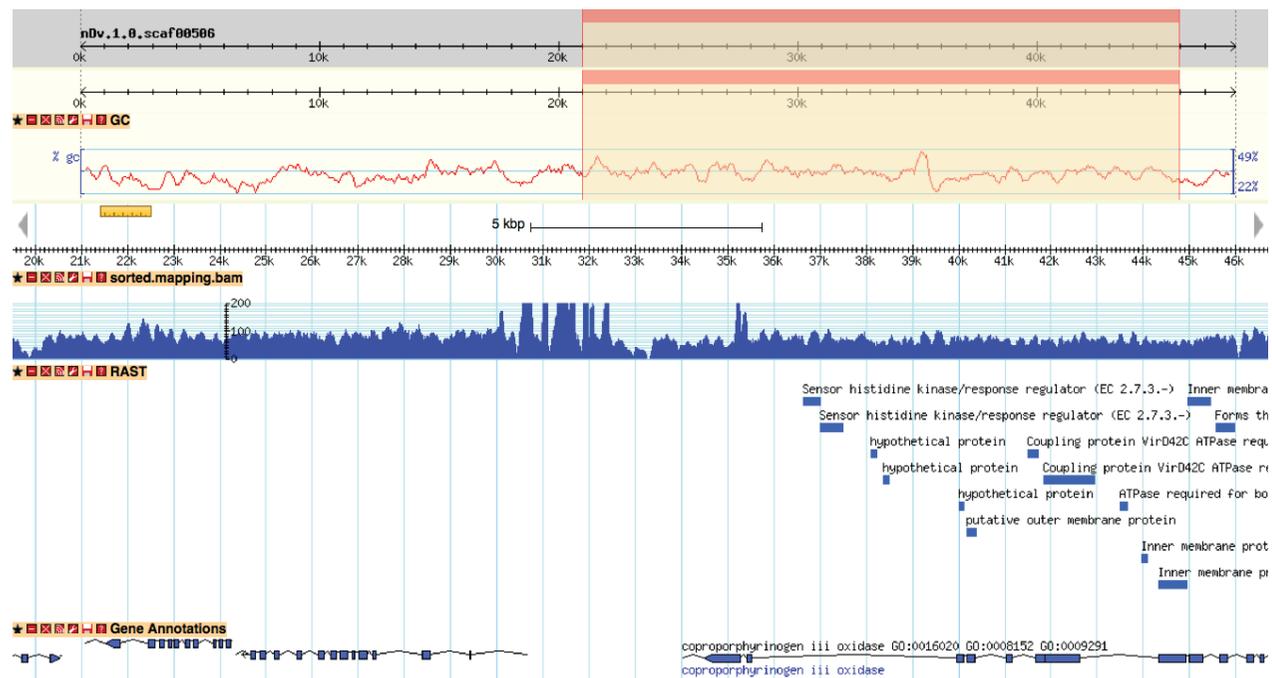

**Figure 4 Comparison of *Wolbachia*-like insertions from two *Dictyocaulus viviparus* isolates, and relationships of the Cameroon *D. viviparus***

**A** 16S rRNA gene fragments from the Cameroon isolate of *D. viviparus* (obtained through whole genome sequencing) and from the Moredun isolate (from specific amplification) are shown aligned. The genome sequence assembly has three copies of *Wolbachia*-like 16S genes, two tandemly arranged and truncated in scaffold scaf09320, and one in scaffold scaf01523.

**B** *ftsZ* gene fragments from the Cameroon isolate of *D. viviparus* (obtained through whole genome sequencing) and from the Moredun isolate (from specific amplification) are shown aligned. While we were able to amplify the complete fragment from the Moredun strain, the genome assembly contains only a truncated *ftsZ* gene (and no consensus is shown for the ~200 bases of essentially unaligned sequence at the 5' end of the alignment).

**C** Bayesian phylogenetic analysis of the complete nuclear small subunit ribosomal RNA (nSSU) genes of the Cameroon *D. viviparus* and other *Dictyocaulus* sp., and outgroups (taken from the European Nucleotide Archive). The Cameroon *D. viviparus* is most similar to the European *D. viviparus* sequenced previously. RAxML analyses yielded the same topology. The 5' gene fragment isolated and sequenced from the Moredun strain was identical to the other *D. viviparus* nSSU sequences.

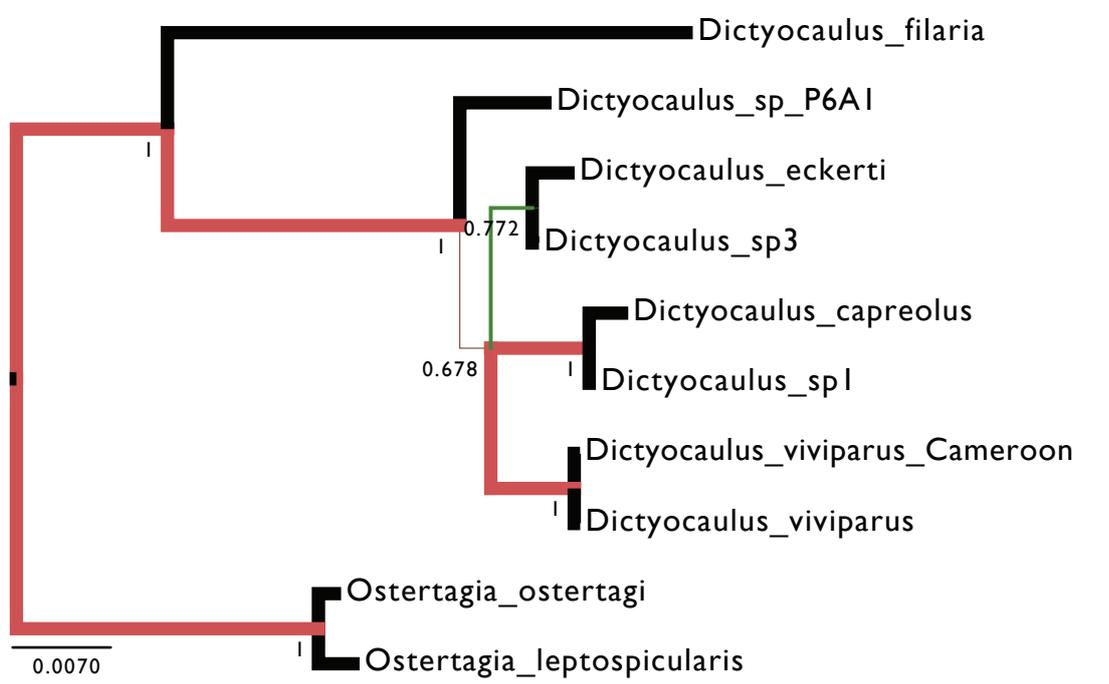

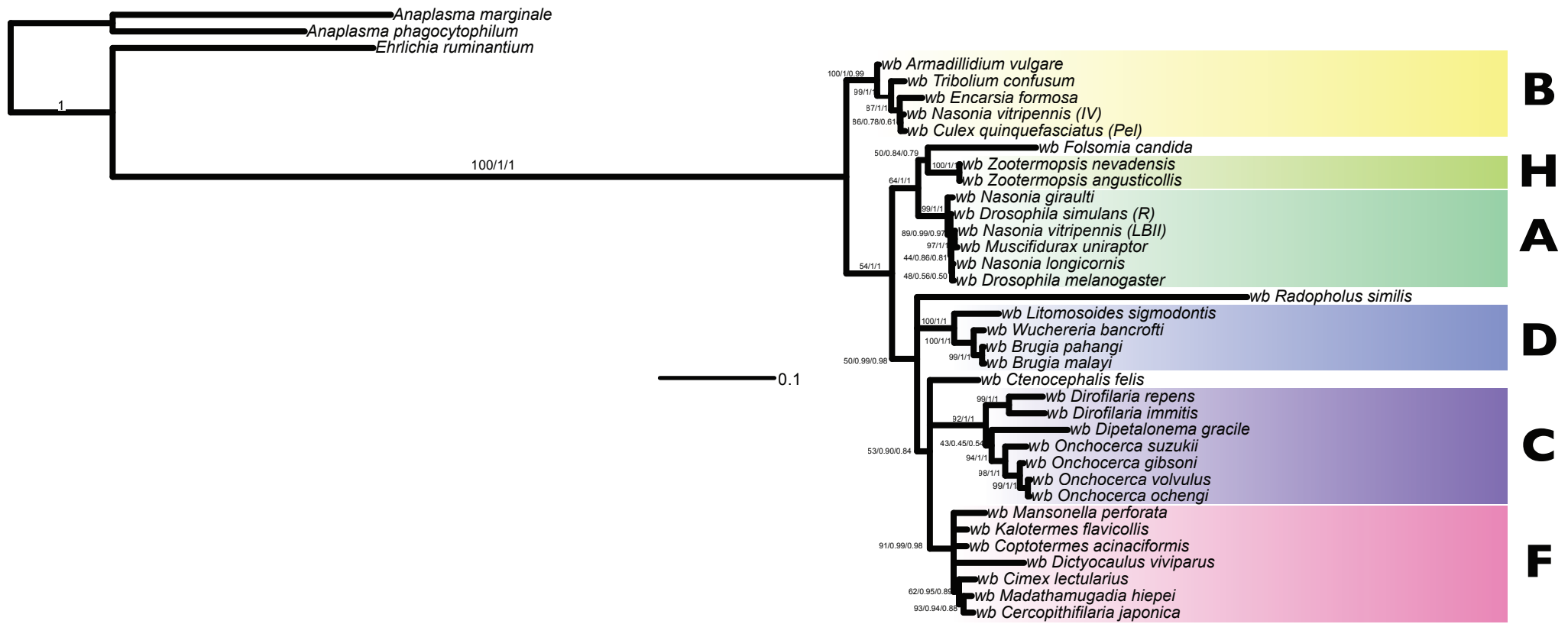

Figure 5 Analysis of the phylogenetic relationships of the Wolbachia nuclear insertions in the Dictyocaulus viviparus genome
Phylogenetic tree inferred from 16S rDNA, groEL, ftsZ, dnaA and coxA loci with maximum likelihood (RAxML) and Bayesian (MrBayes, PhyloBayes) inference. Branch support is reported as (RaxML/MrBayes/PhyloBayes). Strains representing Wolbachia supergroups A, B, C, D, F and H are indicated.